\documentclass[12pt]{article}

\usepackage{latexsym,amsfonts,amsmath,amssymb,pstricks,wasysym,stmaryrd,enumerate,epsfig}
\usepackage{mathrsfs}
\usepackage{theorem}
\usepackage{cite}
\numberwithin{equation}{section}

\newcommand{\be}{\begin{equation}}
\newcommand{\ee}{\end{equation}}
\newcommand{\beu}{\begin{equation*}}
\newcommand{\eeu}{\end{equation*}}
\newcommand{\bea}{\begin{eqnarray}}
\newcommand{\eea}{\end{eqnarray}}
\newcommand{\beaa}{\begin{eqnarray*}}
\newcommand{\eeaa}{\end{eqnarray*}}
\newcommand{\bmx}{\begin{pmatrix}}
\newcommand{\emx}{\end{pmatrix}}

\newcommand{\del}{\partial}

\newcommand{\half}{\frac{1}{2}}

\newcommand{\nn}{\nonumber}

\newcommand{\8}{{\infty}}

\newcommand{\eps}{\epsilon}

\newcommand{\ad}{{\rm ad}}

\newcommand{\ket}[1]{{\,\left|#1\right>}\,}

\newcommand{\ip}{{\cdot}}

\newcommand{\R}{{\mathcal R}}
\newcommand{\id}{{\mathrm{id}}}
\newcommand{\End}{{\mathrm{ End}}}
\newcommand{\A}{{\mathsf A}}

\newcommand{\kP}{{\mathcal U_\kappa (\mathsf P)}}

\newcommand{\tauk}{\tau^{(\kappa)}}

\advance\textheight by \topskip
\begin{document}

\baselineskip 17.5pt
\parindent 18pt
\parskip 8pt

\begin{flushright}
\break

DCPT-08/41

\end{flushright}
\vspace{1cm}
\begin{center}
{\LARGE {\bf  On $\kappa$-deformation and triangular}}

{\LARGE {\bf  quasibialgebra structure}}\\[4mm]
\vspace{1.5cm}
{\large  C. A. S. Young\footnote{\texttt{charlesyoung@cantab.net}} and R. Zegers\footnote{\texttt{robin.zegers@durham.ac.uk}}}
\\
\vspace{10mm}

{ \emph{Department of Mathematical Sciences\\ University of Durham\\
South Road, Durham DH1 3LE, UK}}

\end{center}

\vskip 1in
\centerline{\small\bf ABSTRACT}
\centerline{
\parbox[t]{5in}{\small We show that, up to terms of order $\kappa^{-5}$, the $\kappa$-deformed Poincar\'e algebra can be endowed with a triangular quasibialgebra structure. The universal R~matrix and coassociator are given explicitly to the first few orders. In the context of $\kappa$-deformed quantum field theory, we argue that this structure, assuming it exists to all orders, ensures that states of any number of identical particles, in any representation, can be defined in a $\kappa$-covariant fashion.}}

\vspace{1cm}

\newpage

\section{Introduction}
The $\kappa$-Poincar\'e Hopf algebra $\kP$ \cite{kPoin,kPoind, bicross} is a deformation of (the universal enveloping algebra of) the Poincar\'e algebra, with the strength of the deformation being governed by a parameter $\kappa$ with units of mass \cite{Luk06,LNdsrVk}.

This paper is a continuation of earlier work \cite{YZ,YZ2}, following \cite{DLW1}, whose aim is systematically to construct $\kappa$-deformed quantum field theory from the following particular perspective. (Other approaches can be found in \cite{FKGN,DIKN05,LRZ,AAC,KLM,GW,AAstar,FS94,DJMTWW,KRY,Moller,KMLS98,KMS,GGHMM, GMSS,DMT04}.) We recall the viewpoint on quantum field theory taken by Weinberg in \cite{Weinberg}, namely that quantum field theory takes the form it does because this is essentially the only way to construct a quantum mechanical theory of point particles with Poincar\'e symmetry -- given only a very limited number of additional physical principles, like cluster decomposition. Thus, if one wishes to understand $\kappa$-deformed QFT, it is natural to try to follow this path as closely as possible, making only those modifications forced upon one by the $\kappa$-deformation.

In this approach one begins with particles and scattering theory. The first task is to understand the structure of asymptotic scattering states -- that is, ``in'' states, ``out'' states, or states of a free theory.
Single-particle states transform, by definition, in projective irreducible representations of the Poincar\'e algebra.\footnote{More precisely: single particles transform in projective irreducible representations whose states have \emph{only one continuous label}, which specifies the momentum of the particle.} Poincar\'e and $\kappa$-Poincar\'e are in fact known to be isomorphic as algebras \cite{KLMS94,KLM01,GGM96}, so they share the same representations. Single particle states are consequently well-understood. States of many particles are constructed by taking tensor products of single-particle states. Recall that in order to specify the action of a symmetry algebra on tensor products one requires a coalgebra structure: if
\be \rho_1 : \A \rightarrow \End(V_1), \qquad \rho_2 : \A \rightarrow \End( V_2) \ee
are two representations of an algebra $\A$ then the tensor product $V_1 \otimes V_2$ carries the representation
\be \left(\rho_1 \otimes \rho_2\right) \circ \Delta : \A \rightarrow \End(V_1) \otimes \End(V_2) \cong \End (V_1 \otimes V_2)\ee 
where $\Delta: \A \rightarrow \A \otimes \A$ is the coproduct. As a bialgebra (i.e. an algebra with a compatible coalgebra; see e.g. \cite{Majid}) $\kappa$-Poincar\'e is not isomorphic to Poincar\'e. In particular, the generators of the Poincar\'e algebra obey the usual cocommutative Leibniz rule,
\be \Delta X = X\otimes 1 + 1 \otimes X ,\label{Pcop}\ee
but the coproduct of $\kappa$-Poincar\'e is not cocommutative. This leads to the first major obstacle: in quantum field theory one is concerned with states of many \emph{identical} particles with definite exchange statistics \cite{DLW1,DLW,DLW2,DLW3,Arzano}. For example, the space of states of two bosons of some species transforming in a representation $V$ is usually the quotient of the tensor product $V\otimes V$ by the map $\tau$ which exchanges the two factors:
\be \tau: V\otimes V\rightarrow V\otimes V ;\quad\chi\otimes \psi \mapsto \psi \otimes \chi .\ee
Quotienting out by $\tau$ is a frame-independent operation because $\tau$ is an intertwiner of representations, i.e. it commutes with the action of the Poincar\'e algebra. This in turn holds by virtue of the cocommutativity of (\ref{Pcop}). Cocommutativity means, of course, that $\Delta = \Delta^\text{op}$ where $\Delta^\text{op} := \sigma \circ \Delta$ and $\sigma$ is the flip map
\be \sigma : \A \otimes \A\rightarrow \A \otimes \A; \quad a \otimes b \mapsto b \otimes a .\ee

In the $\kappa$-deformed case, to define states of two identical particles the task is thus to find an intertwiner $\tauk$ to play the role of $\tau$. It is also natural to demand that $\tauk \rightarrow \tau$ as $\kappa\rightarrow \8$, so that we recover the usual notion of particle exchange in the undeformed limit.

\subsubsection*{R matrices}
There is a class of bialgebras whose representations necessarily admit intertwiners: those possessing a \emph{quantum universal R-Matrix}, also called a \emph{quasitriangular} structure. Recall that a universal R-matrix for the bialgebra $\A$ is an invertible element
\be \R \in \A \otimes \A\ee
with the property that, for all $X\in \A$,
\be \R \, \Delta X \,\R^{-1} = \Delta^\text{op} X  .\ee
Let us stress that this is of course not the only axiom that $\R$ must obey for $(\A,\R)$ to be a quasitriangular bialgebra -- and we return to this point in section \ref{sec:phi} below -- but it is, nevertheless, sufficient to guarantee the existence of intertwiners. 
For suppose that $\kP$ possesses such an $\R$, with the property that
\be \R = 1\otimes 1 + O\left(\frac{1}{\kappa}\right).\ee
Then, for any representation $\rho:\kP \rightarrow \End( V )$ of $\kappa$-Poincar\'e, the map
\be \tauk = \tau \circ (\rho\otimes \rho)(\R)\quad \in\quad \End(V\otimes V)\label{intR}\ee 
is an intertwiner with the correct $\kappa\rightarrow \8$ limit. (Note that $\tauk$ can be thought of as the representation of the braided R matrix $\check \R =\sigma\circ \R$, which obeys $\left[ \check\R, \Delta X\right] = 0$.)

An important question is therefore whether or not such an $\R$ actually exists for $\kP$. After recalling the precise definition of $\kP$ in section \ref{sec:kP} below, we address this question to the first few orders in $\frac{1}{\kappa}$ in section \ref{sec:R}.
Then in section \ref{sec:phi}, again working perturbatively in $\frac{1}{\kappa}$, we introduce the coassociator $\Phi$ and discuss states of more than two particles. 
Finally, some conclusions and open questions are given in section \ref{sec:conc}.

\section{The Hopf algebra $\kP$}\label{sec:kP}
The $\kappa$-deformed Poincar\'e algebra in general dimension $1+d$ was first given in \cite{kPoind}. Its generators are
\be M_{ij} = - M_{ji}, \quad N_i, \quad P_i, \quad P_0=E, \qquad\qquad i=1,\dots,d \ee
of, respectively, rotations, boosts, and translations in space and time. The non-vanishing commutators are
\be\left [M_{ij}, M_{kl} \right ] = \delta_{i[l} M_{k]j} + \delta_{j[k} M_{l]i} \, ,\qquad \left [M_{ij}, N_k \right ] = \delta_{k[i} N_{j]} \, . \ee
\be 
 \left[ M_{ij}, P_k \right ] = \delta_{k[i} P_{j]} \ee
\be \left [N_i, P_j \right ] = \delta_{ij} \, \kappa \sinh \left (\frac{E}{\kappa} \right ) \, , \qquad  \left [ N_i, E\right ] = P_i \, , \ee
\be\left [ N_i, N_j \right ] = - M_{ij} \cosh \left (\frac{E}{\kappa} \right ) + \frac{1}{4\kappa^2} \left (\vec P \ip \vec P M_{ij} + P_k P_{[i} M_{j]k} \right ) \, .
\ee
The coalgebra is given by
\bea
\Delta E &=& E \otimes 1 + 1 \otimes E \, , \label{coalg1} \\ 
\Delta P_i &=& P_i \otimes e^{\frac{E}{2\kappa}} + e^{-\frac{E}{2\kappa}} \otimes P_i \, , \\
\Delta N_i &=& N_i \otimes e^{\frac{E}{2\kappa}} + e^{-\frac{E}{2\kappa}} \otimes N_i + \frac{1}{2\kappa} \left (P_j \otimes e^{\frac{E}{2\kappa}} M_{ij}  - e^{-\frac{E}{2\kappa}} M_{ij} \otimes P_j \right ) \, ,\\
\Delta M_{ij} &=& M_{ij} \otimes 1+ 1 \otimes M_{ij} \, \label{coalg2}.
\eea
For completeness, the additional structures which make $\kP$ a Hopf algebra (rather than just a bialgebra) are the antipode and counit maps  
\be S(P_\mu) = -P_\mu, \quad S(M_{ij}) = - M_{ij}, \quad S(N_i) = - N_i + \frac{d}{2\kappa} P_i,\ee
\be \eps(M_{ij}) = \eps(N_i) = \eps(P_\mu) = 0.\ee
Note that we shall work with this ``original'' basis of $\kP$ rather than the bicrossproduct basis of \cite{bicross}; the calculations below are very similar in either basis, but the more ``balanced'' form of the coproduct makes the original basis slightly more convenient to work with for our purposes.

\section{A universal R matrix to $O(1/\kappa^6)$}\label{sec:R}
In this section we look for an invertible element $\R\in \kP \otimes \kP$ with the property that
\be \Delta^{\text{op}}(a) = \R \,\Delta(a) \, \R^{-1}\label{RR}\ee 
for all $a\in \kP$. It suffices to demand that (\ref{RR}) hold for the generators \be\{E,P_i,N_i\},\ee because the $M_{ij}$ and all other elements of $\kP$ are generated by these (and $\Delta$ is, of course, a homomorphism of algebras). We will not be able to make any exact statements at finite $\kappa$, but rather are only able to work order by order in the deformation parameter $\kappa^{-1}$. We shall do so up to terms at $O(\kappa^{-6})$.  
Thus, let
\be \R = e^r =\exp\left(\frac{1}{\kappa}r_1 
              + \frac{1}{\kappa^2} r_2 
              + \frac{1}{\kappa^3} r_3 
+ \frac{1}{\kappa^4} r_4 
+ \frac{1}{\kappa^5} r_5 
              \right) + O\left(\frac{1}{\kappa^6}\right) .\label{Rexp}\ee
One has then the expansion
\be \R \,\Delta(a) \,\R^{-1} = e^{\ad(r)}\Delta(a) 
  := \Delta(a) + \left[r,\Delta(a)\right] + \half \left[r,\left[r,\Delta(a)\right]\right] 
        + \dots \label{adRexp}\ee
which proves useful for calculations -- although note that this series, as written, is certainly not the expansion in inverse powers of $\kappa$: in general \emph{all} of the first $n+1$ terms contribute at order $\kappa^{-n}$.

Equation (\ref{RR}) is true at leading order. At order $\kappa^{-1}$ one finds the equations
\bea \left[ r_1, E\vee 1\right] &=& O(\kappa^{-1}) \\
\left[ r_1, P_i \vee 1 \right] &=& P_i \wedge E + O(\kappa^{-1})\\
 \left[ r_1, N_i \vee 1 \right] &=& N_i \wedge E  + P_j \wedge M_{ij} + O(\kappa^{-1})\eea 
where, for the sake of brevity, we have introduced the notation
\be A \vee B = A \otimes B + B \otimes A \ee
\be A \wedge B =  A \otimes B - B \otimes A.\ee
These equations are solved by
\be r_1 = N_k \wedge P_k \ee
which is the classical r-matrix associated to $\kP$, and has been known since the work of \cite{SZak}; see also \cite{HerranzSantander,Dasz}. The solution $r_1$ is unique up to the addition of terms that commute with $E\vee 1, P_i\vee 1, N_i\vee 1$ to leading order; in other words, terms that are classically\footnote{``classical'' in the sense of ``undeformed''} Poincar\'e invariant. In all dimensions other than 1+2 no such terms exist with the correct mass dimension (i.e. $1$) to match the power of the dimensionful deformation parameter $\kappa$ in the expansion (\ref{Rexp}). The 1+2 dimensional case is very special because there does exist such an invariant, obtained by splitting the Casimir $\eps^{\mu\nu\rho} M_{\mu\nu} P_\rho = -2 N_k P_k + \eps^{ij} M_{ij} E$ symmetrically over the tensor product: $\eps^{\mu\nu\rho} M_{\mu\nu} \vee P_\rho$  \cite{Meusburger,CGST3d}.

At the next order, $\kappa^{-2}$, one finds that there are no ``source'' terms and that $r_2$ must obey simply
\be \left[ r_2, E\vee 1 \right] =  O(\kappa^{-1}), \quad \left[r_2, P_i\vee 1\right]= O(\kappa^{-1}),   \quad \left[ r_2, N_i\vee 1\right]= O(\kappa^{-1}). \ee
These equations do have nonzero solutions of mass dimension 2 in all spacetime dimensions, because there always exists the mass Casimir, $E^2-\vec P \ip \vec P+ O(\kappa^{-1})$ of $\kP$. In $1+3$ dimensions, there are further invariants, $\eps^{\mu\nu\rho\sigma}P_\mu M_{\nu\rho} \vee P_\sigma$ and $\eps^{\mu\nu\rho\sigma}P_\mu M_{\nu\rho} \wedge P_\sigma$. However, let us ignore these possibilities and set \be r_2=0.\ee 
More generally, let us place the following extra condition on $\R=e^r$: 
\be\left[\begin{split} r \text{ must be linear in the Lorentz generators $M_{ij}$, $N_i$\quad}\\\text{and indices must be contracted solely with Kronecker $\delta$'s } \end{split}\right]\label{rassum}\ee
which is a convenient way to fix the freedom we would otherwise have to add homogeneous solutions at each order.

To give a rough motivation for this choice, consider how the intertwiner (\ref{intR}) acts on tensor product states $\ket p \otimes \ket q$, or equivalently on two-particle momentum-space wavefunctions
\be \psi(p,q) \ee 
where we suppress spin degrees of freedom for simplicity. Terms in $r$ containing only momenta $P_i,E$ will merely produce overall factors. It is rather the terms linear in Lorentz generators, whose realizations on wavefunctions involve derivatives, that shift the arguments of $\psi$. And following
\cite{YZ} we expect that
\be \psi(p,q) \underset{\tauk}{\longmapsto} \psi( f(q,p), g(q,p) ) = e^{(f-p)\del_p + (g-q)\del_q} \psi(p,q) .\ee
In the end it is possible that one \emph{should} introduce some scalar prefactor here, perhaps for reasons having to do with the way creation/annihilation operators must be combined into quantum fields, c.f. \cite{DLW3}. But, having noted the freedom to add homogeneous solutions, let us restrict ourselves to (\ref{rassum}) in this paper.

At order $\kappa^{-3}$ one finds after some calculation that
\bea \left[ r_3, E\vee 1 \right] &=& O(\kappa^{-1})\\
     24\left [ r_3, P_i \vee 1 \right] &=& - 3 E^3\wedge P_i + 3 P_i E^2 \wedge E \nn\\
                           && {}+2 P_k P_i \wedge P_i E - 2 P_k \wedge EP_iP_k \nn\\
   && {}+2 E\wedge \vec P \ip \vec P P_i + 2 E P_i \wedge \vec P \ip \vec P - 2 E^2 \wedge EP_i + O(\kappa^{-1}) \\
    24\left [ r_3, N_i \vee 1 \right] 
  &=& 3 N_i \wedge E^3 + 9 E^2 M_{ki} \wedge P_k - 3 E^2 P_k \wedge M_{ki} \nn\\
  && {} - 3 E\wedge E^2 N_i - 6 E\wedge E P_k M_{ik} - 2 \vec P \ip \vec N \wedge EP_i \nn\\
  && {} - 2 P_i P_k \wedge E N_k - 2 N_i \wedge E \vec P \ip \vec P - 2 EP_i P_k \wedge N_k \nn\\
  && {} - 2 P_k M_{ki} \wedge E^2 - 4 P_k \wedge E P_k N_i - 2 E \vec P \ip \vec N \wedge P_i\nn\\
  && {} - 2 E^2 \wedge EN_i - 6 \vec P \ip \vec P N_i \wedge E -4 EP_k \wedge E M_{ki}  \nn\\
  && {} + 4 P_j \wedge P_j P_k M_{ik} - 2 P_j P_i \wedge P_k M_{jk} - 4 E\wedge P_i \vec P \ip \vec N \nn\\     
  && {} -4 EP_k \wedge P_i N_k - 2 M_{ki} \wedge P_k \vec P \ip \vec P \nn\\
  && {} + 6 P_i P_k M_{jk} \wedge P_j - 6 \vec P \ip \vec P M_{ki} \wedge P_k  +O(\kappa^{-1});\eea
and further that these equations have a unique solution obeying (\ref{rassum}), which is
\bea 24 r_3 &=&  {} - 3 E^2 N_k \wedge P_k - 3 N_k \wedge E^2 P_k \nn\\
 && {} - 2 P_l M_{kl} \wedge E P_k + 2 N_k \wedge \vec P \ip \vec P P_k \nn\\
 && {} + 6 \vec P \ip \vec P N_k \wedge P_k - 6 EP_l M_{kl} \wedge P_k \nn\\
 && {}- 2 EN_k \wedge EP_k - 4 P_k P_l N_k\wedge P_l . \eea
Beyond this order direct calculations are somewhat laborious. We have written a program in FORM \cite{FORM} to carry them out, and found
\be r_4 =0\ee
\bea720 r_5  &=&\tfrac{75}{8} E^4N_k\wedge P_k + \tfrac{25}{2} E^3 N_k \wedge EP_k + \tfrac{81}{4} E^2N_k \wedge E^2P_k\nn\\
&& {} + \tfrac{25}{2} EN_k \wedge E^3 P_k + \tfrac{75}{8} N_k \wedge E^4P_k \nn\\
&& {} + \tfrac{75}{2} E^3 P_l M_{kl} \wedge P_k + \tfrac{75}{2} E^2 P_l M_{kl} \wedge E P_k \nn\\
&& {} + \tfrac{81}{2} E P_l M_{kl} \wedge E^2 P_k + \tfrac{25}{2} P_l M_{kl} \wedge E^3 P_k \nn\\
&& {} - \tfrac{21}{2} E^2 N_k \wedge \vec P \ip \vec P P_k - 9 E N_k \wedge E \vec P \ip \vec P P_k -\tfrac{45}{2} N_k \wedge E^2 \vec P \ip \vec P P_k\nn\\
&& {} -\tfrac{135}{2} E^2 \vec P \ip \vec P N_k \wedge P_k - 27 E\vec P \ip \vec P N_k \wedge EP_k - \tfrac{63}{2} \vec P \ip \vec P N_k \wedge E^2 P_k \nn\\
&& {} + 45 E^2 P_l \vec P \ip \vec N \wedge P_l + 18 E P_l \vec P \ip \vec N \wedge EP_l + 21 P_l \vec P \ip \vec N \wedge E^2 P_l \nn\\
&& {} -45 E \vec P \ip \vec P P_l M_{kl} \wedge P_k - 9\vec P \ip \vec P P_l M_{kl} \wedge E P_k \nn\\
&& {} -21 E P_l M_{kl} \wedge \vec P \ip \vec P P_k - 9 P_l M_{kl} \wedge E \vec P \ip \vec P P_k  -36 \vec P \ip \vec P P_j \vec P \ip \vec N \wedge P_j\nn\\
&& {} -12 P_l \vec P \ip \vec N \wedge P_l \vec P \ip \vec P  + 45 \vec P \ip \vec P \vec P \ip \vec P N_k \wedge P_k + 30 \vec P \ip \vec P N_k \wedge P_k \vec P \ip \vec P \nn\\
&&{} -12 P_l P_j N_k \wedge P_k P_l P_j + 9 N_k \wedge P_k \vec P \ip \vec P \vec P \ip \vec P  
            \label{finalR}\eea 

Given (\ref{rassum}), this solution to (\ref{RR}) is unique. Observe that $r$ is antisymmetric. This means that $\R$ is \emph{triangular} i.e.
\be \R_{21} = \R^{-1} \ee
where $\R_{21}$ is the R matrix with the tensor factors flipped. Consequently the intertwiners (\ref{intR}) of representations of $\kP$ are involutive:
\be\tau_{(\kappa)} = \tau_{(\kappa)}^{-1} \quad\Rightarrow \quad\tau_{(\kappa)}^2=\id\label{t21}\ee 
and one can speak of bosons and fermions. (Had $\R$ not turned out to be triangular it would be less clear how to match physics at large $\kappa$ to physics in the undeformed case; though see \cite{Goldin}.)

However, as mentioned above, in order for $\kP$ to be a (quasi)triangular bialgebra there are further requirements on $\R$ in addition to (\ref{RR}). Essentially, $\R$ should obey the quantum Yang-Baxter equation, and as we discuss below the R matrix presented here certainly does not do so.\footnote{The 1+2 dimensional case is exceptional, as we saw above, and in fact there does exist, in this dimension only, a quasitriangular bialgebra structure \cite{CGST3d}, i.e. a quantum R matrix obeying the quantum Yang-Baxter equation. It is genuinely braided, i.e. not triangular.}   
But in fact this is no disaster, because there exists a rather more general notion, that of a \emph{quasitriangular quasibialgebra} structure \cite{Drinfeld}, \cite{MackSchomerus,Majid}. It will turn out that (to the first few orders in $\kappa^{-1}$, at least) $\kP$ does possess such a structure, and that this is sufficient for our purposes. We turn to this now. 

\section{The coassociator and quasibialgebra structure}\label{sec:phi} 
Recall \cite{Majid} that $(\A,\R)$ is a quasitriangular bialgebra if, in addition to (\ref{RR}),\footnote{Following the standard notation, this is an equation in $\A \otimes \A \otimes \A$ and, for example, $\R_{13}$ means $\R$ acting in the first and third tensor factors.} 
\be \left(\Delta \otimes \id\right) \R \overset != \R_{13}\R_{23}, \quad   
\left(\id\otimes \Delta\right) \R \overset ! = \R_{13}\R_{12}.\label{qtbi}\ee   
Intuitively speaking, these are statements about ways of manipulating three ``objects'', initially ordered 1,2,3. For example the second equation, which is sometimes written $\R_{1(23)} = \R_{13} \R_{12}$, says that ``interacting'' 1 with 2 and 3 is the same thing as first interacting 1 with 2 and then interacting 1 with 3. The meaning of ``object'' and ``interact'' depends on the context: in our case, the objects are the labels $p,q,\dots$ of individual constituent particles of a tensor product state.\footnote{For brevity, we implicitly include all the quantum numbers in $p$, including any discrete indices corresponding to spin or internal degrees of freedom.} To interact $p$ with $q$ is the first step in the process
\be \ket p\otimes\ket q \quad \underset{(\rho\otimes \rho)\R}\longmapsto \quad\ket{p'} \otimes\ket{q'} \quad\underset \tau \longmapsto\quad \ket{q'} \otimes \ket{p'} \ee
of exchanging the particles according to the prescription (\ref{intR}). Equations (\ref{qtbi}) are then statements about manipulating the labels of states of three or more particles.\footnote{They would ensure that $\tauk_{ij} = \tau_{ij}\circ (\rho_i\otimes \rho_j) \R$ obey the braid relations $\tau^{(\kappa)}_{i,i+1} \tau^{(\kappa)}_{i+1,i+2}\tau^{(\kappa)}_{i,i+1} \overset != \tau^{(\kappa)}_{i+1,i+2} \tau^{(\kappa)}_{i,i+1}\tau^{(\kappa)}_{i+1,i+2}$, which, together with $\tau^{(\kappa)}_{i,i+1}\tau^{(\kappa)}_{i,i+1} = \id$ in (\ref{t21}) would mean that the $\tau^{(\kappa)}_{i,i+1}$ were a realization of the symmetric group.}
It is straightforward to check directly that they fail at first order in $\kappa^{-1}$ for the R matrix in (\ref{finalR}). This is equivalent to the well-known fact that the classical r-matrix $r_1 = N_k \wedge P_k$ of $\kP$ does not obey the classical Yang-Baxter equation but rather only the modified classical Yang-Baxter equation (MCYBE) \cite{SZak}. 

Fortunately, there is a natural way in which the intuition above about manipulating a string of objects can fail. Suppose that when specifying a state it is necessary to give not only the order of the tensor factors but also a complete bracketing of them: 
\bea \ket{p\,\,(q\,\,r)} &:=& \ket p \otimes \Big( \ket q \otimes \ket r \Big),\nn\\
   \ket{ p\,\, ( (q\,\, r)\,\, s)}\label{brac} &:=& \ket p \otimes\bigg( \Big( \ket q \otimes \ket r \Big) \otimes \ket s \bigg) \eea
and so on. 
A new operation is then needed to move the brackets around. This idea is made precise with the notion of a quasitriangular quasibialgebra structure \cite{Drinfeld, Majid} in which the axioms (\ref{qtbi}) are generalized to
\be \left ( \Delta \otimes \id \right ) \R  = \Phi_{312} \R_{13} \Phi_{132}^{-1} \R_{23} \Phi \, , \qquad \left ( \id \otimes \Delta \right ) \R = \Phi_{231}^{-1} \R_{13}\Phi_{213} \R_{12} \Phi^{-1} \,  \ee
where $\Phi \in \A \otimes \A \otimes \A$ is the \emph{coassociator}. It is required to be invertible, to obey the \emph{pentagon equation} (or \emph{3-cocycle condition})
\be (\id \otimes \id \otimes \Delta) \Phi \,\, (\Delta\otimes \id \otimes \id) \Phi
 = ( 1 \otimes \Phi) \,\, ( \id \otimes \Delta \otimes \id ) \,\, (\Phi \otimes 1) \ee
-- which says that the two ways to perform the rebracketing $((pq)r)s \rightarrow p(q(rs))$ agree -- and to be such that for all $a\in \A$
\be (\id \otimes \Delta) \circ \Delta(a)  = \Phi\,\, \left( ( \Delta \otimes \id) \circ \Delta(a) \right) \,\,\Phi^{-1} .\ee 
In general quasitriangular quasibialgebras can be non-coassociative, but they certainly do not need to be. When coassociativity, i.e. $(\id \otimes \Delta) \circ \Delta=  ( \Delta \otimes \id)\circ \Delta$, holds, as it does for $\kP$, the final condition says simply that $\Phi$ should be invariant, in the sense that
\be \left[\,\, \Phi\,,\,  \left( ( \Delta \otimes \id) \Delta(a) \right) \,\,\right] = 0 \quad \forall a\in \A. \ee

Let us, then, ask whether a $\Phi$ obeying all these conditions exists for $\kP$ and the R matrix of the previous section. It turns out that at least to low orders the answer is yes. We set
\be \nn\Phi =e^\phi= \exp\left( \frac{1}{\kappa} \phi_1
            +\frac{1}{\kappa^2} \phi_2 + \frac{1}{\kappa^3} \phi_3
            +\frac{1}{\kappa^4} \phi_4 \right)
 + O\left(\frac{1}{\kappa^5}\right) \ee
and find by direct calculation a unique solution: 
\bea \phi_1&=&0\label{Phii}\\
    \phi_2 &=&\frac{1}{12} \left( M_{kl} \wedge P_k \wedge P_l 
      + 2N_k \wedge E \wedge P_k\right) \nn\\
\phi_3 &=& \frac{1}{12}\left(1 + \sigma_{13} \right) \Big(
  N_k \otimes EP_k \otimes E + N_k \otimes E \otimes EP_k
- EP_k \otimes N_k \otimes E \nn\\
&& \qquad\qquad\qquad{}   + EN_k \otimes P_k \otimes E - EN_k \otimes E \otimes P_k 
      - P_k \otimes EN_k \otimes E   \nn\\
&&  \qquad\qquad\qquad{}+ P_k M_{lk} \otimes P_l \otimes E - P_k M_{lk} \otimes E \otimes P_l - 
      P_l \otimes P_k M_{lk} \otimes E  \nn\\
&&  \qquad\qquad\qquad{}+ M_{lk} \otimes EP_l \otimes P_k - M_{lk} \otimes P_k \otimes EP_l 
          + P_k \otimes M_{lk} \otimes EP_l  \nn\\
&&  \qquad\qquad\qquad{}+ P_l N_k \otimes P_l \otimes P_k - P_l N_k \otimes P_k \otimes P_l
        + P_l \otimes P_l N_k \otimes P_k \nn\\
&&  \qquad\qquad\qquad{}-  P_k \otimes \vec P \ip \vec N \otimes P_k 
   -2 N_k \otimes P_k P_l \otimes P_l \nn\\
&&  \qquad\qquad\qquad{}+ N_k \otimes \vec P \ip \vec P \otimes P_k + \vec P\ip\vec P \otimes N_k \otimes P_k
 - N_k \otimes P_k \otimes \vec P \ip \vec P    \Big)  \nn   \eea  
where by $\sigma_{13}$ we mean the map $a\otimes b \otimes c \mapsto c \otimes b \otimes a$. We have verified using FORM that $\phi_4$ exists and is unique; its actual expression is very lengthy and we omit it. Note that, as expected, the first non-vanishing term is nothing but $\phi_2=\frac{1}{12}M_{\mu\nu} \wedge P^\mu \wedge P^\nu$, which is the (classically-Poincar\'e invariant) source term that appears in the MCYBE obeyed by the classical r-matrix \cite{SZak,Dasz}. 

\subsubsection*{The role of the coassociator}
At first sight, the need to specify a bracketing of particles within state vectors as in (\ref{brac}) seems very odd, and one might worry that it would introduce a large unwanted redundancy in the space of states. It is important to stress that this is not the case. The important property which must be maintained is that \emph{the counting of states at large $\kappa$ should agree with the counting of states in the usual undeformed case}. That is, there should be a bijection between states at large $\kappa$ and states in the undeformed case. Now the need to write brackets in kets does not spoil this property, so long as we have the means (provided by $\Phi$) to rebracket at will and we declare that states which are related by rebracketings are physically indistinguishable. 

More precisely, define the space of states of $N$ identical particles to be the space of fully bracketed $N$-fold tensor products of single-particle states, quotiented by all exchange and rebracketing operations. Observe that this definition is valid both in the $\kappa$-deformed and undeformed cases: all that is modified are the exchange and rebracketing operations themselves. In the undeformed case rebracketing is trivial (i.e. $\Phi=1\otimes 1\otimes 1$) so to quotient by it is simply to ignore the brackets, and thus the usual definition is recovered.

To give a concrete illustration, consider the simplest example in which rebracketing is possible: the case of three scalar particles. The leading order effect is at order $\kappa^{-2}$. By definition
\be \rho^{\otimes 3}\left(\Phi\right) \,\, \ket{ \left( r \,\, s\right)\,\, t } = \ket{ r' \,\,(s' \,\, t')} \ee
where in view of (\ref{Phii})
\be r'_\nu = r_\nu + \frac{r^\mu t_\mu s_\nu - r^\mu s_\mu t_\nu}{6\kappa^2}+ O\left(\frac{1}{\kappa^3}\right),\quad   
            s'_\nu =  s_\nu + \frac{s^\mu r_\mu t_\nu - s^\mu t_\mu r_\nu}{6\kappa^2} + O\left(\frac{1}{\kappa^3}\right), \nn\ee
\be       t'_\nu=    t_\nu + \frac{t^\mu s_\mu r_\nu - t^\mu r_\mu s_\nu}{6\kappa^2}+ O\left(\frac{1}{\kappa^3}\right). \label{rst}\ee

We can now make contact with the perturbative results for states of three scalar particles of mass $m$ (transforming in $V_m$) given in the appendix of \cite{YZ2}. It was shown there that to $O(\kappa^{-3}$) there exists a one-parameter family of pairs of maps 
\be \tau_1, \tau_2 : V_m \otimes V_m \otimes V_m \rightarrow  V_m \otimes V_m \otimes V_m \ee
such that 
\be \tau_1^2=\tau_2^2=\id, \quad\quad \tau_1\,\tau_2\,\tau_1 = \tau_2\,\tau_1\,\tau_2 \ee
(i.e. $\tau_1, \tau_2$ realize the symmetric group $S_3$) and that, in the limit $\kappa\rightarrow \8$, $\tau_1$ exchanges the first and second tensor factors, $\tau_2$ the second and third. For a certain choice of the parameter ($a=0$) one has
\be \tau_1 = \tau^{(\kappa)} \otimes 1 \equiv \tau^{(\kappa)}_{12} \ee
where it may be verified that $\tau^{(\kappa)}$ is indeed the intertwiner (\ref{intR}) obtained by representing the R matrix found in section \ref{sec:R}.\footnote{Note that the appendix to \cite{YZ2} and the entirety of \cite{YZ} used the bicrossproduct basis, so it is necessary to translate to that basis using the relations in \cite{bicross} to check this match.} But then it turns out that $\tau_2 \neq \tau^{(\kappa)}_{23}\equiv1\otimes\tau^{(\kappa)}$. This is now as expected: $\tau_2$ should instead be
\bea \tau_2 &=&\rho^{\otimes 3}\left(\Phi^{-1}\right) \circ \tau_{23}^{(\kappa)} \circ \rho^{\otimes 3}\left(\Phi\right) \nn\\
&=&\rho^{\otimes 3}\left(\Phi^{-1}\right) \circ (1 \otimes \tau) \circ \rho^{\otimes 3}\left(1\otimes \R\right)\circ \rho^{\otimes 3}\left(\Phi\right) \eea
because this is the operation which, starting from a state bracketed as
\be \ket{ (r\,\,s) \,\,t}  \ee
first moves the bracket, \emph{then} exchanges the second and third particles, and then returns the bracket to its initial ``reference'' position. And one can see that the $r,s,t$ terms in $\tau_2$ at $a=0$ in \cite{YZ2} correspond to (\ref{rst}). Note that taking another value of the parameter $a$ for $\tau_1$,$\tau_2$ would correspond to choosing a different linear combination of the bracketings $\ket{ (r\,\,s) \,\,t}$  and  $\ket{ r\,\,(s \,\,t)}$ as the reference configuration to which the state is returned after each flip operation. 

This scheme extends to states of $N$ particles in a natural fashion. To each choice of reference bracketing, for example
\be \ket{ (\dots ((( p \, \, q) \,\, r \,\, )\,\, s\,\, ) \quad\dots\quad )\,\, t},\ee
is associated a realization of the symmetric group $S_N$. For the particular choice above one has
\bea \tau_1 &=& \tauk_{12} \nn\\
     \tau_2 &=&  \rho^{\otimes N}(\Phi^{-1}_{123})\circ \tauk_{23} \circ \rho^{\otimes N}(\Phi_{123})   \nn\\
     \tau_3 &=&  \rho^{\otimes N}(\Phi^{-1}_{(12)34})\circ \tauk_{34} \circ \rho^{\otimes N}(\Phi_{(12)34})   \nn\\
   &\vdots&\nn\\
  \tau_N &=&  \rho^{\otimes N}(\Phi^{-1}_{(1\dots N-2)\,\,N-1\,\,N})\circ \tauk_{N-1\,\,N} \circ \rho^{\otimes N}(\Phi_{(1\dots N-2)\,\,N-1\,\,N})\eea
where the definition
\be \Phi_{(1\dots k)\,\,k+1\,\,k+2} = (\Delta^{k-1}\otimes \id\otimes \id)\Phi \ee
is unambiguous by coassociativity of $\kP$. 

We should stress though that the definition of the space of states of $N$ identical particles itself, as given above, is independent of any choice of preferred bracketing. 

\section{Conclusions and outlook}\label{sec:conc}
In this paper we have shown that the $\kappa$-deformed Poincar\'e Hopf algebra $\kP$ possesses a triangular quasibialgebra structure up to terms of order $\kappa^{-5}$. 
It appears likely, given the rather intricate way in which the relevant algebraic equations turned out to be soluble, that this structure persists to all orders in $\kappa^{-1}$. The obvious open problem is to give a proof of existence to all orders. 

Since $\mathcal U_\kappa(\mathsf P_{1,3})$ is a contraction limit of $\mathcal U_q(\mathsf{so}(2,3))$ one approach might be to ask whether the triangular quasibialgebra structure of $\kP$ is inherited from $\mathcal U_q(\mathsf{so}(n))$. 
It cannot be the limit of the standard \emph{quasi}triangular structure on $\mathcal U_q(\mathsf{so}(n))$, whose R matrix indeed diverges in the limit (except in the special case of dimension three \cite{CGST3d}). But  $\mathcal U_q(\mathsf{so}(n))$ can also be endowed with the structure of a triangular quasibialgebra \cite{MajidBegs} and it would be interesting to see whether there is a limit of this reproducing the $\R$ and $\Phi$ above. It would also be very nice to have a more geometrical understanding of the intertwiners of representations, perhaps in the spirit of \cite{FKG}; this might be another way to obtain exact rather than perturbative results.

One thing to note is that although it is conceptually valuable to check that the triangular quasibialgebra structure is exact, in practice knowledge of its explicit form at higher orders in $\kappa^{-1}$ seems unlikely to be important for physics. If $\kappa$ is finite in nature, it is certainly very large. Moreover it is usually supposed \cite{KG-S,FKGS,ASS} -- though cf. \cite{Meusburger} -- that the role of quantum field theory with $\kappa$-deformed Poincar\'e symmetry, if any, will be that of an effective description in a regime intermediate between standard QFT and full quantum gravity.

To restate the central point of this work: what the existence of the triangular quasibialgebra structure ensures is that there is a fully $\kappa$-covariant way to define states of many identical particles (in any representation) in such a way that these states are in bijection with the states of the theory in the undeformed case. If $\kappa$ is to be large but finite in (an effective theory of) the real world, the fact that this property holds is crucial: if it did not, the counting of states would be affected in ways we should have already observed. 


\vspace{2cm}
{\it Acknowledgements.} \,\,C.Y. is grateful to L. Freidel, J. Lukierski, S. Meljanac and S. Majid for interesting discussions and suggestions. C.Y. is funded by the Leverhulme trust. R.Z. is funded by an EPSRC postdoctoral fellowship.   

\bibliographystyle{unsrt}
\bibliography{Rmat.bib}

\end{document}